\documentclass[a4paper]{jpconf}
\usepackage{graphicx}
\usepackage{amsmath} 
\usepackage{verbatim}

\begin{document}
\title{Active power control of wind farms: an instantaneous approach on waked conditions}

\author{Jean Gonzalez Silva$^1$, Bart Matthijs Doekemeijer$^2$, Riccardo Ferrari$^1$, Jan-Willem van Wingerden$^1$.}

\address{$^1$ Delft University of Technology, Delft, 2628CD The Netherlands}
\address{$^2$ National Renewable Energy Laboratory, Golden, CO 80401, United States}

\ead{\{J.GonzalezSilva, r.ferrari, J.W.vanWingerden\}@tudelft.nl and bart.doekemeijer@nrel.gov}

\begin{abstract}
This paper presents a closed-loop controller for wind farms to provide active power control services using a high-fidelity computational fluid dynamics based wind plant simulator. The proposed design enhances power tracking stability and allows for simple understanding, where
each turbine is considered as a pure time-delay system.
The paper investigates the control performance with different nominal power distributions in a fully waked condition and limited power availability.
Results demonstrate the improvement in power production obtained by closing the control loop, compared to greedy operation. Additionally, power tracking capabilities are enhanced 
with a nominal power distribution favored by axial-induction, as well as the occurrence of turbine saturation and the distribution of loads.
\end{abstract}

\section{Introduction}
\hspace{10pt} Electrical grid stability is an important yet unresolved issue in the current transition from fossil-fuel-based to renewable energy sources. The demand for wind turbines to provide ancillary grid services is growing significantly, with legislation being put in place in several countries \cite{aho2012}. This so-called active power control (APC) comes in various forms, depending on the timescale of the required power response. 

Herein, a wind farm controller that enforces power reference tracking assigned by a transmission system operator (TSO) is designed. This type of active power control is classified as automatic generation control or frequency regulation in the literature \cite{vanWingerden2017}. The literature on APC for individual wind turbines is more widespread, but outside of the scope of this paper. 
Examples of wind turbine APC algorithms can be found in  Aho et al. \cite{aho2016} and Kim et al. \cite{kim2018}. 

In 2002, Rodriguez-Amenedo et al. \cite{amenedo2002} presented an initial simulation study on APC in wind farms. In that study, wake effects are implicitly included in an open-loop manner by feeding the simulation with field measurements of the ambient wind speed corresponding to a real wind farm with wake effects. Moreover, the work includes a supervisory control system that dispatches additional power demand signals to derated turbines in the situation that other turbines reach their maximum power production. While tested in a low-fidelity model and the limited validation study, it provided an excellent starting point for APC in wind farms. Succeeding publications such as Hansen et al. \cite{hansen2006}, Spudic et al. \cite{spudic2010, spudic2012, spudic2015},
Biegel et al. \cite{biegel2013}, Badihi et al. \cite{badihi2015}, Zhao et al. \cite{zhao2015}, Madjidian \cite{madjidian2016}, Ahmadyar and Verbic \cite{ahmadyar2017}, Siniscalchi-Minna et al. \cite{siniscalchi2018} and Bay et al. \cite{bay2018} entail a mixture of open-loop and closed-loop APC algorithms that often enforce a secondary objective in addition to power tracking, typically being the minimization of a structural load measured on turbines (e.g., \cite{spudic2010, spudic2012, spudic2015, madjidian2016}) 
or maximization of the available power (e.g., \cite{ahmadyar2017, siniscalchi2018}). The algorithms presented in these publications vary from simple PI-based controllers \cite{spudic2012} to more complicated algorithms such as distributed model predictive control \cite{spudic2015, bay2018} and adaptive pole placement \cite{badihi2015}. However, many algorithms in the literature are tested in situations where sufficient power was available in the wind (e.g., \cite{hansen2006, badihi2015, madjidian2016, siniscalchi2018}), which do not account for realistic scenarios of low power availability. Additionally, all of these algorithms were tested in low-fidelity simulation environments, and their applicability in practice remains uncertain.

Fleming et al. \cite{fleming2016} is the first publication, to the best of our knowledge, to perform a large-eddy simulation study for APC. The wind-farm-wide power reference signal is equally dispatched among the turbines, despite the fact that an inequal amount of power is available at each turbine due to wake formation. However, the proposed controller in that work does not include feedback, which leads to situations in which individual turbines cannot track their power reference signal. This result highlights the need for a controller that actively dispatches the demanded power signal among turbines through feedback and highlights the importance of including heavy-wake-loss situations in validation studies. Since its publication, an increasing number of APC algorithms was tested in large-eddy simulation. Notable works are the model predictive controllers of Boersma et al. \cite{boersma2019}, Shapiro et al. \cite{shapiro2017} and Vali et al. \cite{vali2021}. The main issue with all these controllers is their complexity, which is often a large barrier for adoption.

Accordingly, van Wingerden et al. \cite{vanWingerden2017} presented a gain-scheduled PID controller that dispatches a correction signal on the turbine power demand to track a wind-farm-wide power demand. 
The rotor forces are therein described using actuator line models (ALMs) \cite{churchfield2017,troldborg2008}. 
Vali et al. \cite{vali2019} extended this work by adding load minimization of the tower base bending moment as a secondary control objective. The algorithm was successfully tested in a large-eddy simulation environment under heavy-waked conditions, but with simplistic static actuator disk models for the turbine rotors. Then, Silva et al. \cite{silva2021} extend on that work by also adding a secondary control loop which balances the aerodynamic loads on the turbines, while exploring its benefits on turbine saturation scenarios using ALMs. 

In the current work, a wind farm controller that enforces power reference tracking is designed by considering the turbines as pure time-delay systems and validated through Large-Eddy Simulations (LES) in a waked situation. The simulation environment used is the Simulator fOr Wind Farm Applications (SOWFA) developed by the U.S. National Renewable Energy Laboratory \cite{churchfield2012}. 
DTU 10MW reference turbines \cite{bak2013} represented by ALMs are resolved. 
Being under low power availability, the individual available power in the wind might be lower than the individual demanded power, leading to the so-called \emph{turbine saturation}. As a direct extension of Wingerden et al. \cite{vanWingerden2017}, the contributions of this work are
\begin{enumerate}
	\item the complete closed-loop APC algorithm for wind farms, which is simple to understand and to implement compared to the
	literature;
	\item the wind turbine controller with guaranteed stability constraint;
	\item excellent wind turbine and wind farm tracking performance, limited by the sampling time of the simulation (or in practice, of the real system);
	\item a high-fidelity simulation study  with presence of \emph{turbine saturation} is performed to validate the control solution.
\end{enumerate}

The remainder of the paper is structured as follows. Section \ref{mm} outlines the synthesis of the proposed power-tracking wind turbine controller and the proposed supervisory wind farm controller. Section \ref{res} reports a high-fidelity simulation study to validate the controllers. Section \ref{conc} concludes the paper.

\section{Methods} \label{mm}

\subsection{Synthesizing a power-tracking wind turbine controller} \label{powertrackingmethod}

\hspace{10pt} The wind turbine controller is synthesized to track a reference power signal, whenever possible. The controller presented here shows resemblance with the pitch-reserve controller described in Fleming et al. \cite{fleming2016} and the KNU2 algorithm in Kim et al. \cite{kim2018}. 
These controllers present reduction of axial induction as turbine is derated. This allows more energy from the wind to pass for the downstream turbines and lower aerodynamic loads, which benefits the farm operation.

\begin{figure}[h]
	\begin{minipage}{18pc}
		\includegraphics[width=21pc]{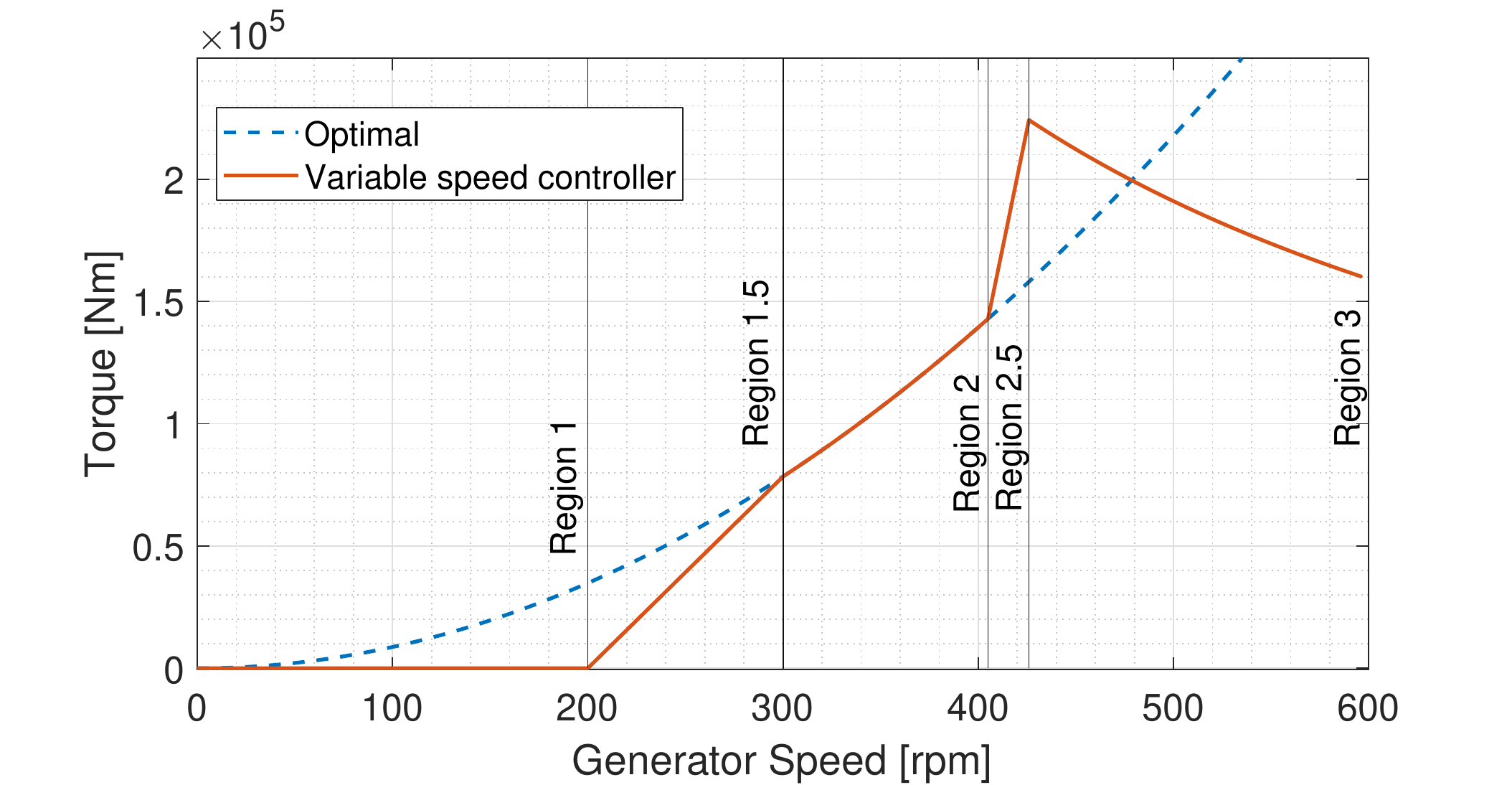}
		\caption{\label{greedytorque} \small Generator torque control law of {\it greedy control}.}
	\end{minipage}\hspace{2pc}%
	\begin{minipage}{18pc}
		\includegraphics[width=19pc]{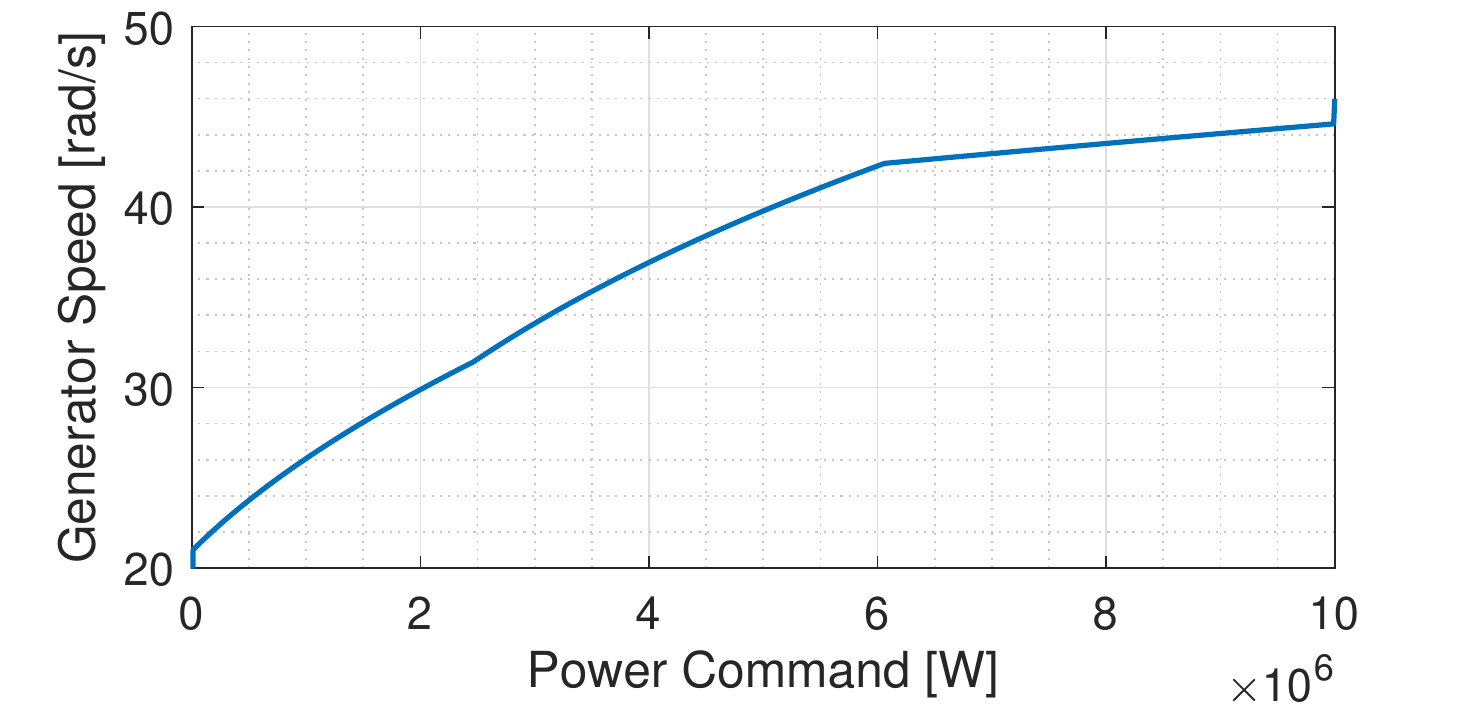}
		\caption{\label{rotorreference} \small Generator speed reference as a function of the demanded power.}
	\end{minipage} 
\end{figure}

In the proposed controller, power tracking is achieved via a gain-scheduled PID pitch control by reducing the rotor speed setpoint as a function of the demanded power. As a feedback loop, the pitch controller seeks to regulate the rotor speed to the desired reference speed.   
This PID controller is similar to the controller used for rotor speed regulation in traditional region III but the rotor speed setpoint is tabulated as a look-up table depicted in Figure \ref{rotorreference} rather than a rated value. The literature standard for wind turbine control is referred to as {\it greedy control}. The reader is referred to \cite{pao2009,jonkman2009} for more information on {\it greedy control} and \cite{mulders2018,rosco2021} for a practical implementation of such a controller for the 10MW turbine.
The rotor speed setpoint is then selected by the same amount of power that would be generated by the {\it greedy control}, where the look-up table is obtained through the generator torque-speed curve in Figure \ref{greedytorque}.

Accordingly, a generator torque control law is herein implemented to avoid undesirable behaviours that lead to turbine shutdown. In this work, two operational modes for the generator torque control can be distinguished.

\subsubsection{Control mode I: forced power reference tracking} \label{prtrack} \hspace*{\fill}

The generator torque $\tau_{gen,\, \, tracking}$ necessary to meet a certain power demand can be expressed as
\begin{equation} \label{torquelawtracking}
\tau_\mathrm{gen, \, \, tracking} =  P_\mathrm{dem} (\omega_\mathrm{gen} \eta_\mathrm{gen})^{-1},
\end{equation}
where $P_\mathrm{dem}$ is the demanded generator power, $\omega_\mathrm{gen}$ is the generator speed, $\tau_\mathrm{gen}$ is the generator torque, and $\eta_\mathrm{gen}$ is the generator efficiency. This control law only requires the measurements of $\omega_\mathrm{gen}$ and achieves a theoretical perfect power tracking in the absence of time delays in the system. Due to the high inertia associated to the rotor, $\omega_\mathrm{gen}$ changes slowly and thus near-perfect tracking can be achieved even with time delays. The generator efficiency $\eta_\mathrm{gen}$ is assumed to be a constant of which the value is known a priori.

In the situation of $P_\mathrm{dem}$ reduction, the rotor will spin up with the torque law in Equation \eqref{torquelawtracking}. As a consequence, the rotor speed is regulated by the pitch control, which pitches the blades according to the $P_\mathrm{dem}$, allowing the controller to converge to a new equilibrium. 

Moreover, in the situation of $P_\mathrm{dem}$ increase, the rotor will slow down with the torque law in Equation \eqref{torquelawtracking}. This can be particularly problematic, for two reasons. First, both the rotor speed and the aerodynamic efficiency\footnote{The power coefficient $C_\mathrm{P}$ is the ratio of power extracted compared to the available power in the wind.} $C_\mathrm{P}$ will tend to reduce and an increasingly higher generator torque becomes necessary to meet the desired power production. The second reason is that the turbine operation can get close to the stall region as a result of the reduction of the tip speed ratio\footnote{The tip speed ratio $\lambda$ is the ratio of the linear speed of the blade tip compared to the inflow wind speed.} $\lambda$. In this region, because of the complete flow separation, the turbine loses the ability to generate aerodynamic torque. Consequently, this leads to a drastic and undesirable reduction of the rotor speed. These behaviours must be avoided as they lead to a turbine shutdown.
To prevent them, a secondary control mode is introduced in the next subsection.

\subsubsection{Control mode II: greedy control and power reference tracking} \label{control2}
\hspace*{\fill}

In the {\it greedy control}, the generator torque  is stable for $\omega_\mathrm{gen}\geq 0$, and globally converges to the optimal power coefficient $C_\mathrm{P}$ to maximize the turbine's power production (in region II). The greedy torque control law\footnote{In the industry, the optimal torque law in region II is often replaced with a PID-controller-based tip-speed-ratio tracking algorithm in combination with a wind speed estimator. Such an algorithm does not sufficiently add to the relevance of this work and therefore is outside the scope of this paper.}, which is composed of distinct regions, is represented by $\tau_\mathrm{gen, \, \, greedy}$ and depicted in Figure \ref{greedytorque}. Combining the greedy torque with the power tracking control law yields
\begin{equation} \label{torquelawcombined}
\tau_\mathrm{gen, \, \, combined} = \textrm{min} \left( \, \tau_\mathrm{gen, \, \, greedy}, \, \, \tau_\mathrm{gen, \, \, tracking} \,\right).
\end{equation}

The generator torque control law in Equation \eqref{torquelawcombined} ensures that the turbine does not operate at a lower tip-speed ratio than expected due to fast transients. As a result,  imminent shutdowns are prevented and the turbine operation remains farther from stall being the approach more conservative. 
Thus, we address the instability issues described in Section \ref{prtrack}. 
Therefore, a theoretical perfect power tracking is achieved whenever  $\tau_\mathrm{gen, \, \, tracking}$ is not constrained by $\tau_\mathrm{gen, \, \, greedy}$ and $P_\mathrm{dem} \leq P_\mathrm{greedy}$, where $P_\mathrm{greedy}$ is the hypothetical power produced by the greedy control with the current wind inflow.

As it is not always possible that the inequality $P_\mathrm{dem} \leq P_\mathrm{greedy}$ holds due to low available power in the wind, the turbine controller switches to {\it greedy control } whenever the collective blade pitch angle reaches the switch value $\theta_\mathrm{switch}$ and the generator speed becomes lower than the reference speed. This keeps the turbine producing the maximum power possible at greedy conditions, while the turbine is saturated.


\subsection{Synthesizing a wind farm controller} \label{wfcontrol}
\hspace{10pt} When wind farms rather than single wind turbines are to track a reference power signal, a power setpoint distribution over the turbines must be decided upon. The power that a turbine can produce is directly correlated to the local wind speed and varies within the farm due to wake interactions. As consequence, 
distributing a wind-farm-wide power reference signal over individual turbines is a non-trivial problem. This paper follows van Wingerden et al. \cite{vanWingerden2017} to synthesize a model-free and closed-loop controller to distribute the power setpoints among the turbines and minimize the wind-farm-wide reference tracking error, with several significant simplifications to promote understanding and improved tracking performance.

The input signal of a single turbine is the demanded power $P_\mathrm{dem}$, and the actual power produced $P_\mathrm{gen}$ is an output. In the situation that the turbine saturation does not occur, i.e. $P_\mathrm{dem} \leq P_\mathrm{greedy}$, and at near-perfect tracking, the input-output relationship is
\begin{equation}
P_\mathrm{gen}^k = \tau_\mathrm{gen, \, \, tracking}^k \,\, \omega_\mathrm{gen}^k \,\, \eta_\mathrm{gen} = P_\mathrm{dem}^{k-1} \,\, (\omega_\mathrm{gen}^{k-1} \,\, \eta_\mathrm{gen})^{-1}  \,\,\omega_\mathrm{gen}^k \,\, \eta_\mathrm{gen},
\end{equation}
where $k$ is the discrete time index of the simulation and controller. With a sufficiently high sampling rate, we can assume $\omega_\mathrm{gen}^k \approx \omega_\mathrm{gen}^{k-1}$ and therefore, $P_\mathrm{gen}^k \approx P_\mathrm{dem}^{k-1}$. Thus, the wind turbines can be considered as near-perfect pure time-delay systems\footnote{Pure time-delay systems have their response delayed by a time period. The result  $P_\mathrm{gen}^k \approx P_\mathrm{dem}^{k-1}$ should be verified by the adopted power tracking method at individual turbines. 
This is investigated in Section \ref{indpowertrackingsection}.} with their time delay equal to the simulation sampling time $\Delta t$, where the time $t^k=t^{k-1}+\Delta t$. 
Time-delay systems inherently limits controller design due to right-half-plane zero (nonminimum-phase) behavior.

Now, consider power tracking at the farm scale. The wind-farm-wide reference $r^k$ is to be divided
among the turbines. Mathematically, the demanded power signal for each turbine $P^k_\mathrm{dem, \,i}$ is
\begin{equation}
P^k_{\mathrm{dem, \,} i} = \alpha_i r^k + \Delta u^k, \, \, \textrm{ with } \sum_{i=1}^{N_\mathrm{T}} \alpha_i = 1 .
\end{equation}

The term $\alpha_i$ divides the total wind farm power over the turbines, defined as the nominal active power distribution, here assumed to be time invariant for simplicity.\footnote{The derivation can straightforwardly be extended to include time dependency in $\alpha_i$, as in Silva et al. \cite{silva2021} by an additional closed-loop with the thrust forces. Also, time-varying distributions can be designed to consider constrained load turbines due to detected faults and failures.} The term $\Delta u^k$ accounts for turbine saturation ($P^k_{\mathrm{dem, \,} i} > P^k_{\mathrm{greedy, \,}i}$). 
The correction term $\Delta u^k$ is the output of a pure integrator controller based on van Wingerden et al. \cite{vanWingerden2017}, here defined as
\begin{equation}
u^k = u^{k-1} + K_\mathrm{I} e^k \Delta t, \, \, \textrm{ with } e^k = r^k - \bar{P}^k,
\end{equation}
where $K_\mathrm{I}$ is the integrator gain. The pure integrator controller is designed to eliminate the instantaneous wind-farm-wide tracking error $e^k$ from the wind-farm-wide reference $r^k$ to the sum of all individual active power production $\bar{P}^k$. Then the integrator gain is chosen as $K_\mathrm{I} = N_\mathrm{T}^{-1}\Delta t^{-1}$, where $N_\mathrm{T}$ is the number of turbines in the farm, which is by definition the optimal controller for a time-delay system.\footnote{Gain-scheduling due to turbine saturation is not implemented to avoid undesirable significant transients in the demanded power observed at small farms. Therefore, the wind farm control operates sub-optimally when turbines saturate.} The error $e^k$ would be therefore eliminated on the next time-step whenever not all turbines are saturated.

Using the integrator, the wind farm power tracking stability is assured, where the accumulated error tends to be eliminated in steady-state by incrementing the active power variation $\Delta u^k$ on each nominal active power reference $\alpha_i r^k$. 
Integrator anti-windup is implemented when all turbines are saturated. Moreover, the integrator state resets whenever all turbines are not saturated. This controller achieves near-perfect tracking limited by the time delay and overall power availability.

\subsection{Overview of controller}

\begin{figure}[h!]
	\centering
	\includegraphics[width=\linewidth]{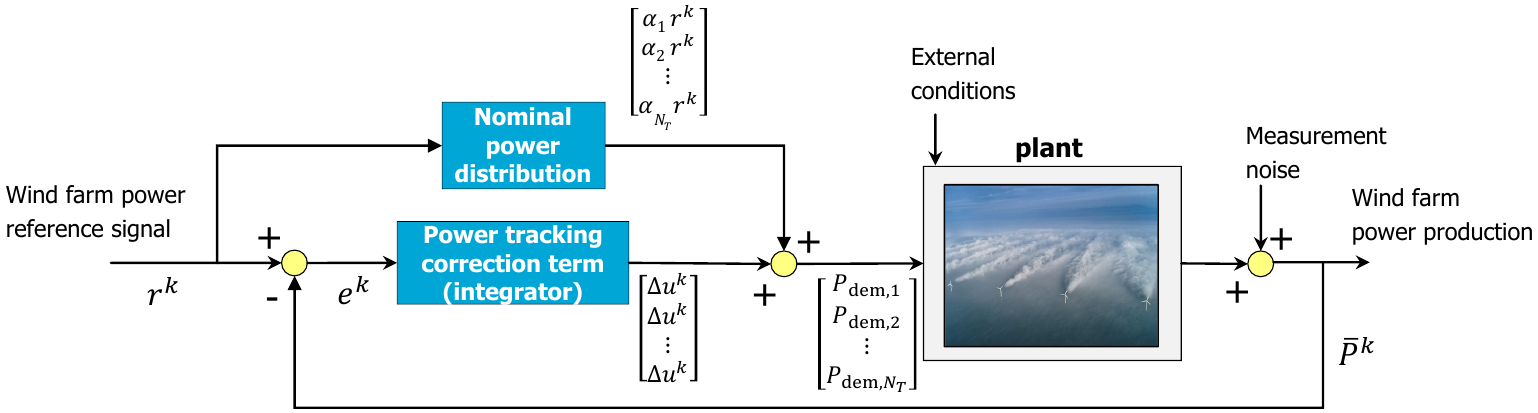}
	\caption{\small Controller framework: the feedforward control loop ensures power tracking when turbines are not saturated. The feedback loop enables power tracking when one but not all turbines are saturated by integrating the wind farm power tracking error and dividing it across the non-saturated turbines.}
	\label{overview}
\end{figure}

\hspace{10pt} An overview of the wind farm controller architecture is shown in Figure \ref{overview}. This figure shows the feedforward loop being the nominal power distribution with $\alpha_1 ... \alpha_{N_\mathrm{T}}$, and the feedback loop
accounting for the saturated wind turbines through $\Delta u^k$. Furthermore, the controller parameters are summarized in Table \ref{contr_param}. The tuning parameters for the individual wind turbine controller are derived from the respective literature. Additionally, the wind farm controller parameters are found and described in Section \ref{wfcontrol}.
\begin{table}[h]
	\caption{\label{contr_param} Controller parameters} 
	\small
	\begin{center}
		\lineup
		\begin{tabular}{*{3}{l}}
			\br                              
			Variable&Symbol&Value\cr 
			\mr
			Proportional gain of the pitch controller & - & Gain-scheduled; 0.039 - 1.41 s \cite{rosco2021}\cr
			Integral gain of the pitch controller & - & Gain-scheduled; 0.067 - 0.28 s \cite{rosco2021} \cr
			Derivative gain of the pitch controller & - & 0.0 s \cite{rosco2021}\cr 
			Corner frequency of generator speed low-pass & - & 0.1798 Hz \cite{rosco2021}\cr
			filter & & \cr
			Generator efficiency & $\eta_\mathrm{gen}$ & 1 \cite{mulders2018, rosco2021} \cr
			Generator torque constant for greedy control & $K_\mathrm{greedy}$ & 79.43986 N-m/(rad/s)$^2$ \cite{rosco2021} \cr
			Transitional generator speed bet & - & 200.0 rpm \cite{rosco2021} \cr
			Transitional generator speed - region 1.5 to 2 & - & 300.0 rpm \cite{rosco2021} \cr
			Transitional generator speed - region 2 to 2.5 & - & 405.0 rpm \cite{rosco2021} \cr
			Rated generator slip percentage in region 2.5 & - & 	10.0 \cite{jonkman2009} \cr 
			Rated power & - & 10 MW \cite{bak2013} \cr
			Transitional generator speed between regions   & - & 95.0 \cite{jonkman2009} \cr
			2.5 and 3 percentage of rated generator speed & & \cr
			Rated generator speed & - & 445.67 rpm \cite{jonkman2009} \cr
			Maximum generator rate & - & 15,000 N-m/s \cite{mulders2018, rosco2021} \cr
			Maximum blade pitch rate & - & 10 deg/s \cite{mulders2018, rosco2021} \cr
			Fine blade pitch angle & $\theta_\mathrm{fine}$  & 0.75 deg \cite{rosco2021} \cr
			Switch blade pitch angle &$\theta_\mathrm{switch}$  & 1 deg \cite{rosco2021} \cr 
			Integral gain of the wind farm controller &$K_\mathrm{I}$  & $N_T^{-1}\Delta t^{-1}$ \cr
			
			\br
		\end{tabular}
	\end{center}
\end{table}
\section{Results}\label{res}
\subsection{Simulation setup}
\hspace{10pt} 
The parameters for the SOWFA simulations is set as shown in Table \ref{sowfapar}.  First, the power tracking on an individual turbine is investigated to be considered as a pure time-delay system. Then, a small 3-turbines-farm as illustrated in Figure \ref{sowfaflow} is simulated with fully overlapping wakes. A baseline open-loop simulation, three simulations with different closed-loop controller configurations, and one greedy simulation in open-loop are looked into. 
Essentially, these simulations assess the difference in where in the farm the turbines are derated, looking at power
reference tracking, turbine saturation, and loads. 
\begin{table}[h]
	\caption{SOWFA simulation parameters}
	\small
	\label{sowfapar}
	\begin{center}
		\begin{tabular}{l c}
			\hline
			Property  & Value \\
			\hline
			Sub-grid-scale (SGS) model & One-equation eddy viscosity\\
			Domain size & 3 km $\times$ 3 km $\times$ 1 km\\
			Cell size outer regions & 10 m $\times$ 10 m $\times$ 10 m\\
			Cell size near rotor & 2.5 m $\times$ 2.5 m $\times$ 2.5 m\\
			Simulation time-step & 0.1 s \\
			Atmospheric boundary layer (ABL) stability & Neutral \\
			Mean inflow wind speed & 9 m/s\\
			Surface roughness & 0.15 m \\
			Turbulence intensity & 0.0 and 5.0 \% \\
			Turbine rotor approximation & Actuator Line Model (ALM) \\
			Turbine type & DTU 10 MW \cite{bak2013} \\
			Turbine rotor diameter & 178.3 m \\
			Turbine hub height & 119 m \\
			Blade smearing factor & 5.0 m \cite{troldborg2008} \\
			Inter-turbine spacing & 5 D \\
			\hline
		\end{tabular}
	\end{center}
\end{table}



\subsection{Power tracking performance at a wind turbine level} \label{indpowertrackingsection}
\hspace{10pt} The power demand is forced with a step to verify the ability to respond as a pure time-delay system using a single turbine.
\begin{figure}
	\begin{center}
		\includegraphics[width=\linewidth]{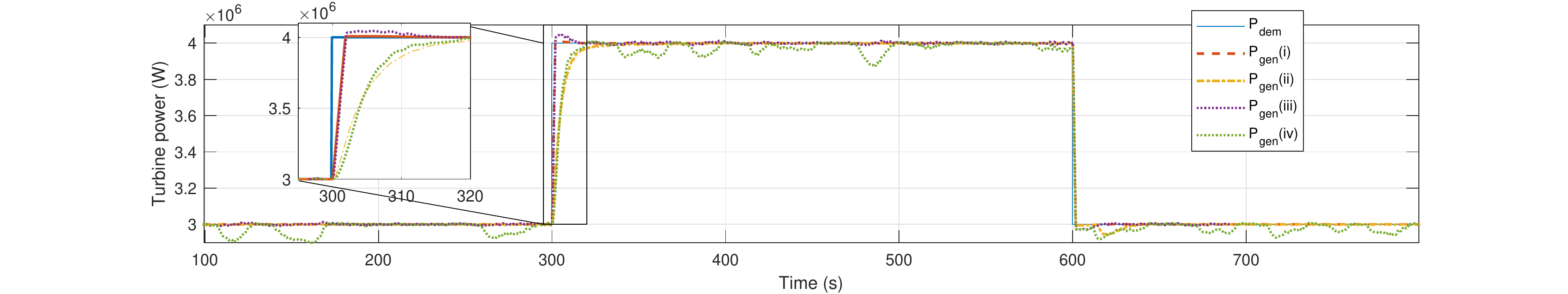}
	\end{center}
	\caption{\label{powertrackind} \small Power response of demanded power steps: (i) control mode I with TI=0\%; (ii) control mode II with TI=0\%; (iii) control mode I with TI=5\%; (iv) control mode II with TI=5\%.}
\end{figure}
\begin{figure}[h]
	\begin{minipage}{18pc}
     \includegraphics[width=16pc]{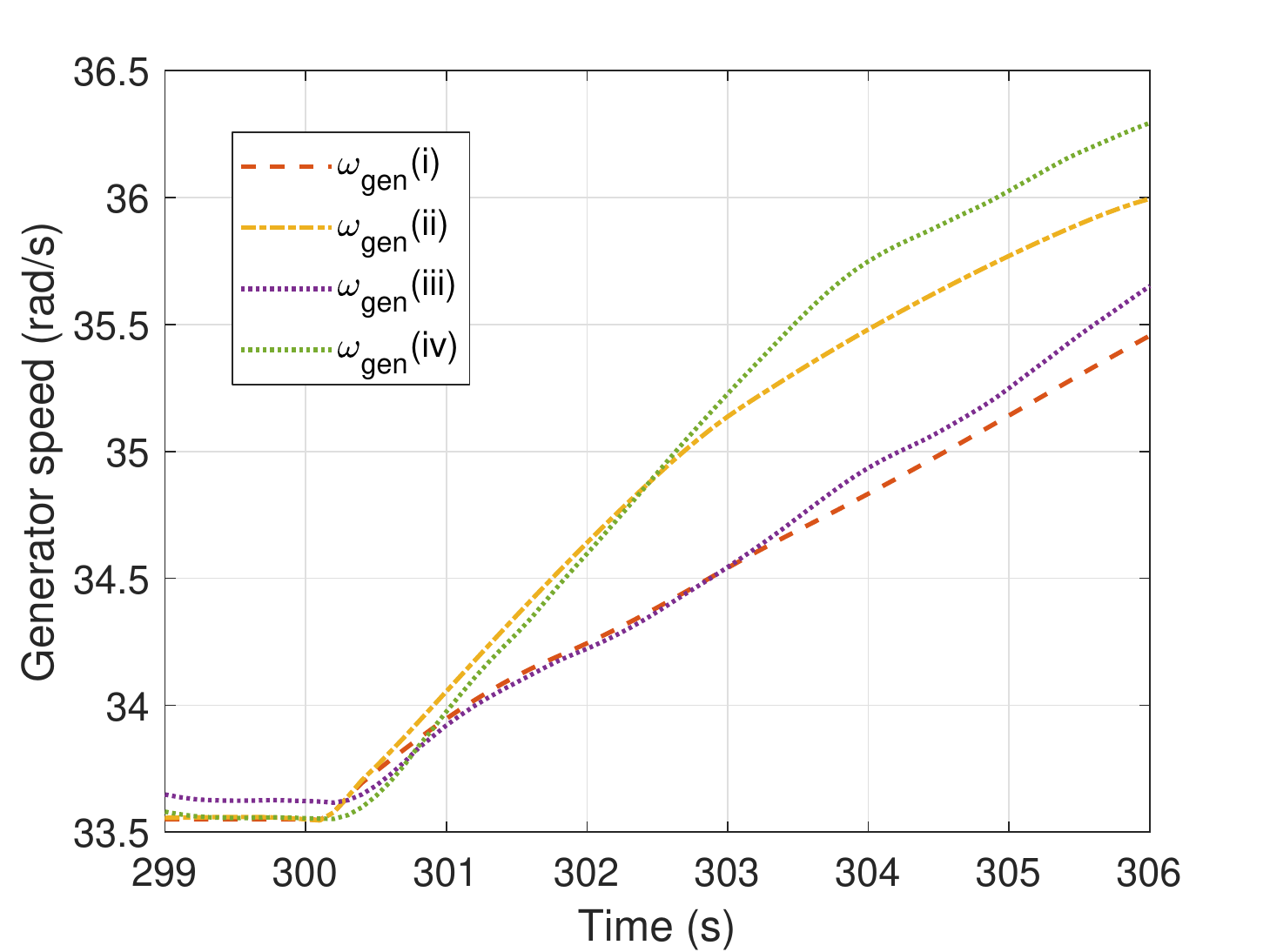}
		\caption{\label{genspeed} \small Generator speed response of demanded power steps: (i) control mode I with TI=0\%; (ii) control mode II with TI=0\%; (iii) control mode I with TI=5\%; (iv) control mode II with TI=5\%. }
		
	\end{minipage}\hspace{1.5pc}%
	\begin{minipage}{18pc}
    	\includegraphics[width=18pc]{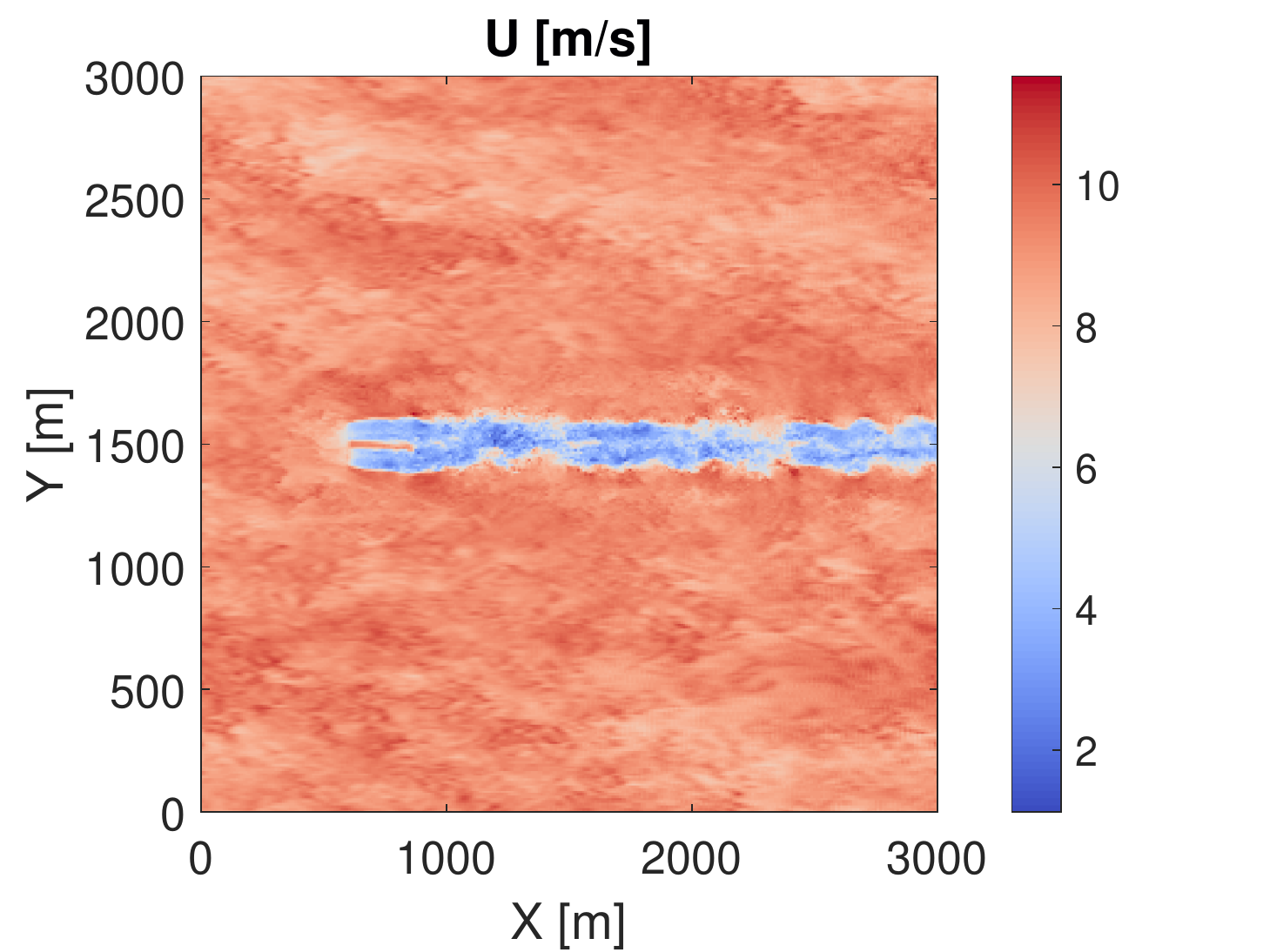}
		\caption{\label{sowfaflow} \small An instantaneous horizontal slice of flow output taken
			from a Simulator fOr Wind Farm Applications (SOWFA)
			simulation.}
	\end{minipage} 
\end{figure}
As shown in Figure \ref{powertrackind}, the power tracking method from Section \ref{powertrackingmethod} has presented a satisfactory response to be considered as a pure time-delay system. At the simulated conditions, 
the response of the control mode II presents a pure time-delay behaviour for positive and negative power demanded variations of 112 kW/s and 592 kW/s, respectively. For the  control mode I, 545 kW/s and 593 kW/s, respectively. In Figure \ref{genspeed}, the corresponding slow changes in the generator speed are depicted.
Therefore, we adopt the control mode II as elucidated in Section \ref{control2} because of its reasonable fast power tracking response.

\subsection{Power tracking performance at the wind farm}
\hspace{10pt} 
To follow a constant $10$ MW wind farm active power reference signal $r^k$, different nominal active power distribution $\alpha_i$ are proposed as listed in Table \ref{farmscenarios}.
In case 0 (baseline), the wind farm is set to individually track an uniform nominal active power distribution in opened-loop, i.e. no feedback is used. Differently, in cases 1, 2 and 3, the closed-loop feedback is implemented with different fixed nominal active power distributions to assess their effects on the farm operation. Lastly, in the greedy case 4 the power production is individually maximized in open-loop.\footnote{The greedy case considers to extract individually the maximum amount of power and was presented superior performance in terms of power maximization than axial-induction based control in \cite{annoni2016}.}

\begin{figure}[h!]
	\centering
	\includegraphics[width=\linewidth]{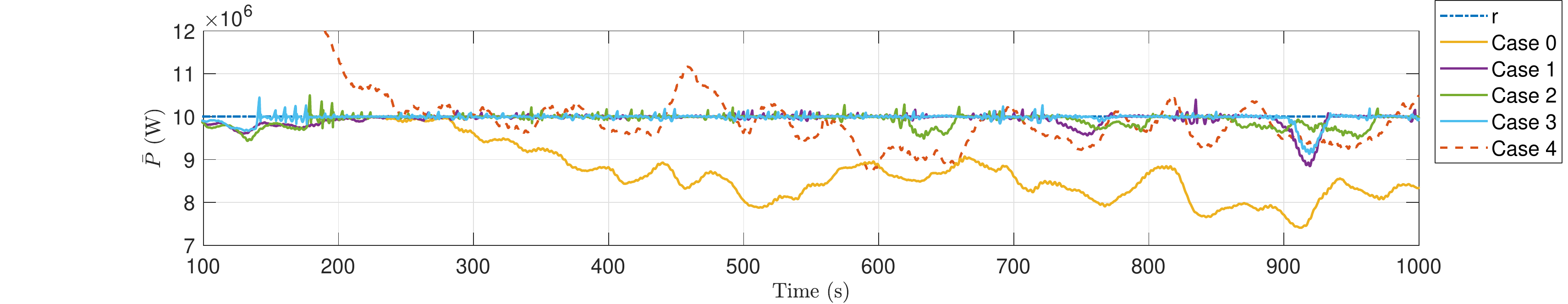}
	\caption{Wind farm active power production. Case 3 presents the best tracking performance following the constant active power reference $r$. In addition, the closed-loop solution shows a better power production than the greedy case 4 in the study.}
	\label{totalpower}
\end{figure}

\begin{figure}[h!]
	\centering
	\includegraphics[width=\linewidth]{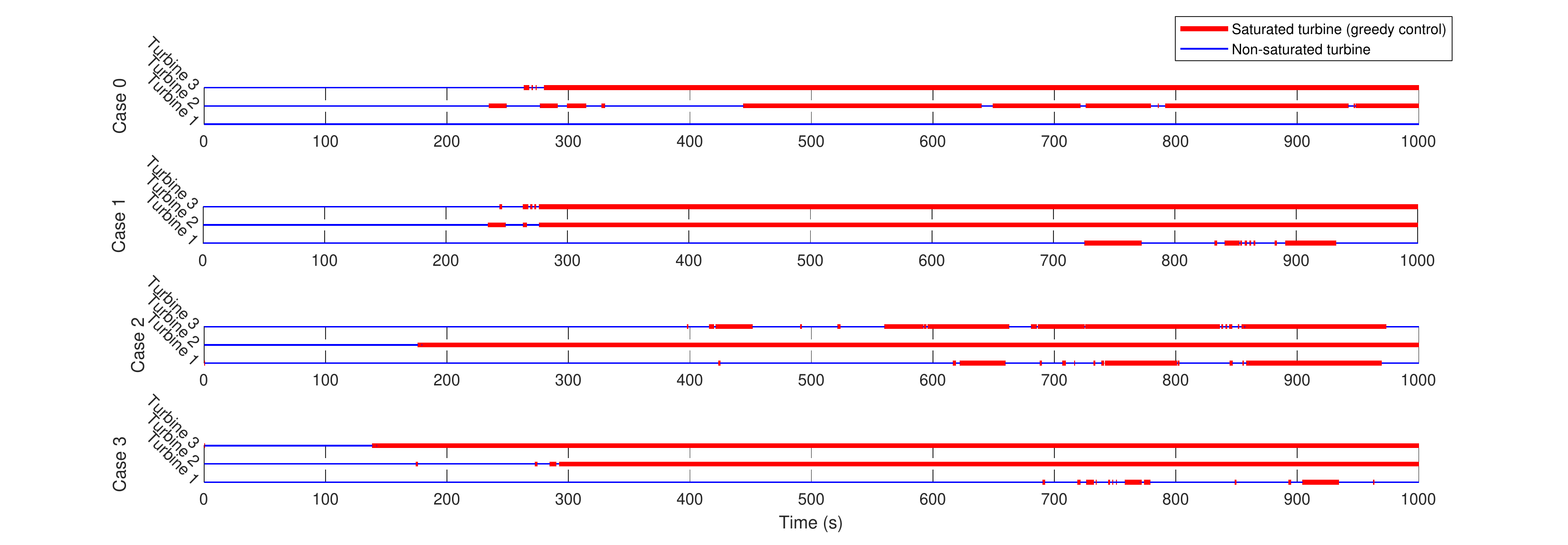}
	\caption{Turbine saturation with distinct pre-fixed nominal power distributions. }
	\label{saturation}
\end{figure}

The results show that in case 0 no compensation leads to the worst active power tracking. On the other hand, the power tracking is mostly kept by using the closed-loop solution.
When upstream turbines are derated in case 3, more energy remains in the wind flow, which is extracted at a later time instant by downstream machines. This behavior also avoids turbine saturation. Therefore, case 3 allows for the best reference tracking while presenting great aerodynamic load distribution. Derating upstream machines reduces the aerodynamic loads of themselves and reduces turbulence in the wake. Although increasing the absolute wind speed impinging downstream machines, having an negative effect on the loads of downstream turbines, it results in a fair load distribution. In contrast, Case 2 presents the worst result among closed-loop cases, in which rapidly occurs full saturation of the farm. The time evolution of the total farm power production $\bar{P}$ is shown in Figure \ref{totalpower}, where the corresponding turbine saturation is in Figure \ref{saturation}.
\begin{table}[h!]
	\caption{Power tracking performance of simulations}
	\small
	\label{farmscenarios}
	\begin{center}
		\begin{tabular}{l c c c c c}
			\hline
			Case & $\alpha_i$ & Mean  & \% & RMS Track & Mean Thrust$_i$\\
			&  & Power (MW)  & Change &  Error (MW) & ($\times 10^5$ N)\\
			\hline
			0 (baseline) & [33.3, 33.3, 33.3] & 8.654 & - & 1.48 & [4.75,    7.44,    5.74] \\
			1 & [33.3, 33.3, 33.3] & 9.955 & +15.04\% & 0.16 & [8.50,    6.50,    6.07] \\
			
			2 & [50.0, 33.3, 16.7] & 9.937 & +14.82\% & 0.14 & [10.33,    5.62,    5.19] \\
			
			3 & [16.7, 33.3, 50.0] & 9.980 & +15.32\% & 0.11 & [8.14,    6.83,    5.99] \\
			
			4 (greedy) & N/A & 9.834  & +13.64\% & 9.85  & [11.70,    5.01,    5.73] \\
			\hline
		\end{tabular}
	\end{center}
\end{table}

\section{Conclusion}\label{conc}

\hspace{10pt} This paper has presented a novel design of the APC for a wind farm considering individual turbines as pure time-delay systems and its validation using a high-fidelity wind plant simulator, SOWFA. The total active power tracking response was improved significantly with the closed-loop solution in the study, reaching up to 15\% of improvement on power production compared with the baseline. 
In addition, the nominal active power distribution yields different performance in closed-loop. Following the axial induction concept, case 3 has presented the best active power tracking and reasonable aerodynamic load distribution. This is an indication that axial induction control is still relevant for the development of closed-loop approaches and future technologies, and hence, more investigation.
The improvement on having closed-loop control is also seen in comparison with greedy case, having a better power production in terms of the active power outcome and aerodynamic loads. Yet undesirable small spikes and oscillations were observed on the active power of the closed-loop solution, so more studies should be carried out to either eliminate or accommodate them. 

Future research will elaborate smart time-varying distribution of the nominal active power by predicting available power, as well as, consider designed constrained turbines due to faults and failures in the proposed APC solution.

\ack{The authors would like to acknowledge the WATEREYE
project (grant no. 851207). This project has received funding from the European Union Horizon 2020 research and innovation programme under the call H2020-LC-SC3-2019-RES-TwoStages.}

\section*{References}


\end{document}